\documentclass[
  reprint,
  superscriptaddress,
  aps, prx,
  longbibliography
]{revtex4-2}

\usepackage{graphicx}
\usepackage{dcolumn}
\usepackage{bm}
\usepackage{amsmath,amssymb,amsfonts}
\usepackage{xcolor}
\usepackage[colorlinks=true,linkcolor=blue,urlcolor=blue,citecolor=blue]{hyperref}
\usepackage{placeins}

\usepackage[normalem]{ulem}

\begin{document}


\title{Real-time Dynamics in 3D for up to 1000 Qubits with Neural Quantum States: Quenches and the Quantum Kibble--Zurek Mechanism}

\author{Vighnesh Dattatraya Naik}
\affiliation{%
 Theoretical Physics III, Center for Electronic Correlations and Magnetism,
 Institute of Physics, University of Augsburg, 86135 Augsburg, Germany}%

\author{Zheng-Hang Sun}
\thanks{Contact author: \href{mailto:zech-email@gmail.com}{zhenghang.sun@uni-a.de}}
\affiliation{%
 Theoretical Physics III, Center for Electronic Correlations and Magnetism,
 Institute of Physics, University of Augsburg, 86135 Augsburg, Germany}

\author{Markus Heyl}
\affiliation{%
 Theoretical Physics III, Center for Electronic Correlations and Magnetism,
 Institute of Physics, University of Augsburg, 86135 Augsburg, Germany}%
\affiliation{%
 Centre for Advanced Analytics and Predictive Sciences (CAAPS),
 University of Augsburg, Universitätsstr.\ 12a, 86159 Augsburg, Germany}

\date{\today}

\begin{abstract}
The exponential complexity of the quantum many-body wave function limits accurate numerical simulations of real-time dynamics, especially for high-dimensional
systems beyond 1D, where rapid entanglement growth poses severe challenges for conventional methods. Neural Quantum States (NQS), a framework based on artificial neural networks, have emerged as a powerful approach for real-time dynamics in 2D, but their scalability and accuracy in 3D have remained an open challenge. Here, we establish NQS as a scalable framework for quantum dynamics in 3D by introducing a residual-based convolutional architecture tailored to cubic spin lattices. Focusing on the 3D transverse-field Ising model, we demonstrate that NQS reliably capture distinct quench regimes, including collapse-and-revival dynamics and, most challengingly, the dynamics following a sudden quench to the quantum critical point. Building on this, we perform finite-rate quenches to the critical point on lattices containing up to $1000$ qubits, an unprecedented system size for numerical simulations of real-time quantum dynamics beyond 1D. This enables the first large-scale numerical demonstration of the 3D quantum Kibble--Zurek mechanism (QKZM). The QKZM in 3D is particularly intriguing because it lies at the upper critical dimension of the Ising universality class ($d+z=4$) where the standard power laws are modified by logarithmic factors together with prominent sub-leading logarithmic corrections. By deriving these corrections from renormalization-group flow equations up to two-loop order, we obtain a robust data collapse across all simulated system sizes for the correlation function, the excess energy, and the quantum Fisher information, the latter revealing universal multipartite-entanglement dynamics. In all cases, we find compelling agreement with the expected scaling dimensions. Our findings establish NQS as a scalable and reliable tool for exploring nonequilibrium phenomena in 3D quantum matter and
for providing numerical benchmarks for quantum simulators realizing 3D models.
\end{abstract}

\maketitle

\section{Introduction}
The real-time dynamics of quantum many-body systems represents a central frontier in modern physics, driven by fundamental questions concerning thermalization~\cite{Polkovnikov2011RMP,DAlessio2016AdvPhys}, universality~\cite{Polkovnikov2011RMP, Dziarmaga2010AdvPhys,Bray1994AdvPhys}, information spreading~\cite{Lieb1972CMP,Calabrese2005JStatMech}, and transport~\cite{Bertini2016PRL,Doyon2020ARCM}.
These regimes are becoming increasingly accessible in state-of-the-art quantum-simulation platforms~\cite{Georgescu2014RMP} such as Rydberg-atom arrays~\cite{Labuhn2016Nature,Bernien2017Nature,Ebadi2021Nature,Gross2017Science}, trapped-ion systems~\cite{Blatt2012NatPhys,Britton2012Nature,Zhang2017Nature}, superconducting qubits~\cite{Houck2012NatPhys,Yan2019Science}, and ultracold quantum gases~\cite{Bloch2008RMP,Bloch2012NatPhys, Greiner2002Nature,Kinoshita2006Nature}.
A quantitative theoretical description of such dynamical quantum matter is therefore not only key for theoretical discoveries and explorations. It is therefore of equal relevance for benchmarking quantum simulators as well as providing predictions and protocols for guiding future experiments~\cite{Georgescu2014RMP,Gross2017Science}.

Numerical simulations, however, face severe challenges due to the exponential complexity of the quantum many-body wavefunction.
While dynamics in one dimension are often accessible using Matrix Product States~\cite{Vidal2004PRL,Schollwock2011AnnPhys}, extending these methods to higher dimensions remains highly challenging.
In 2D, tensor-network approaches such as Projected Entangled Pair States~\cite{Verstraete2004PRL} or Tree Tensor Networks~\cite{Shi2006PRB} can access certain dynamical regimes but are typically limited to short times due to rapid entanglement growth~\cite{Calabrese2005JStatMech, Eisert2015NatPhys}.
In 3D, the computational cost of tensor-network contractions increases even more sharply, rendering large lattices essentially inaccessible for nonequilibrium dynamics~\cite{Vlaar2021PRB, Tindall2025arXiv}. 
A recent approach based on real-time operator evolution via Sparse Pauli Dynamics~\cite{Tomislav2025PRXQuantum} has shown significant performance in 2D, but encounters severe limitations in 3D due to both computational complexity and convergence issues.
Conversely, although Dynamical Mean-Field Theory~\cite{Georges1996RMP,Aoki2014RMP} becomes exact in the limit of infinite dimensions, it is challenged by the complexity in taking into account spatial quantum correlations in the physically relevant 2D and 3D regimes. 
As a result, 3D quantum dynamics remain one of the most challenging frontiers in the numerical simulation of quantum many-body systems.

Neural Quantum States~(NQS) have emerged as a promising alternative for simulating quantum dynamics, demonstrating significant success in 1D and 2D~\cite{Carleo2017Science, Schmitt2020PRL, Schmitt2022SciAdv, SchmittHeyl2025arXiv, Naik2025arxiv}.
While the NQS framework is not inherently constrained by spatial dimensionality, scaling it into a practical, controlled tool for 3D remains an open challenge, primarily due to the increased complexity of optimization and the need for architectures that can efficiently incorporate 3D locality and symmetries.

\begin{figure*}[!t]
\centering
\includegraphics[width=1.0\textwidth]{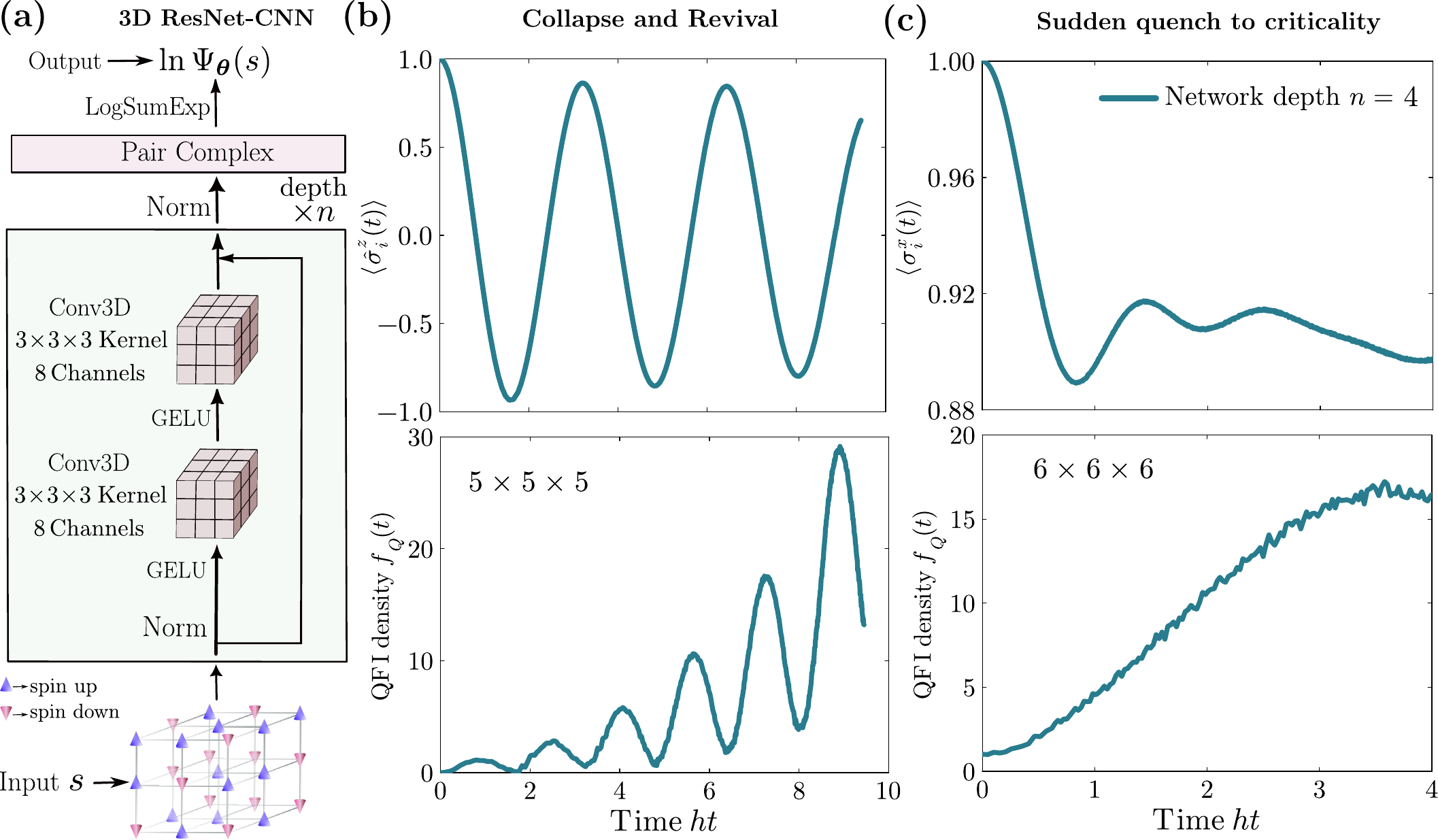}
\caption{\textbf{Neural-network architecture and quench dynamics.}
\textbf{(a)} Schematic of the 3D ResNet-CNN architecture.
\textbf{(b)} Quench from a ferromagnetic product state deep into the paramagnetic regime at $h=2h_c$, leading to collapse-and-revival oscillations of the longitudinal magnetization $\langle\sigma_i^{z}(t)\rangle$ (top) and a stepwise increase of the quantum Fisher information density $f_Q(t)$ (bottom), reflecting the generation of multipartite entanglement.
\textbf{(c)} Quench to the critical point $h=h_c$ starting from the paramagnetic product state, showing the dynamics of  transverse magnetization $\langle\sigma_i^{x}(t)\rangle$ (top) and QFI density $f_Q(t)$ (bottom).}
\label{fig:Figure0}
\end{figure*}

In this work, we establish NQS as a scalable framework for solving numerically exactly 3D quantum dynamics by introducing a convolutional architecture tailored to cubic lattices (see Fig.~\ref{fig:Figure0}(a)), building on the ResNet-CNN Architecture approach established for 2D systems in Ref.~\cite{chen2024nature, chen2025arxiv}.
We provide compelling numerical evidence that our NQS framework is capable to access dynamical regimes in 3D quantum matter beyond what has been achievable with existing methods.
This includes in particular the first numerical demonstration of the quantum Kibble--Zurek mechanism in 3D quantum matter.

As a paradigmatic model, we study the 3D transverse-field Ising model under several nonequilibrium protocols.
We show that NQS can capture collapse-and-revival oscillations~\cite{Greiner2002Nature} together with the buildup of strong multipartite entanglement~\cite{Jurcevic2017PRL,Schmitt2020PRL}.
We further study the dynamics following a sudden quench to the quantum critical point, a particularly challenging regime for numerical simulation, where our NQS results set a new state-of-the-art benchmark by reaching unprecedented timescales~\cite{Tomislav2025PRXQuantum}.

By leveraging NQS on lattices of up to \(1000\) spins, we provide compelling numerical evidence for the 3D quantum Kibble--Zurek mechanism (QKZM)~\cite{Kibble1976JPhysA,Zurek1985Nature,Zurek2005PRL,Dziarmaga2005PRL}.
The 3D quantum Ising model lies at the upper critical dimension~\cite{Li2024PRE, Coester2016PRB} of the Ising universality class, \(d+z=4\), where the leading power-law scaling is modified by logarithmic corrections together with prominent subleading inverse-logarithmic terms.
By incorporating these corrections through the integration of the renormalization-group flow equations up to two-loop order, we establish a refined theoretical benchmark for our numerical results.
Using this framework, we obtain a compelling data collapse for equal-time correlation functions, the excess energy, and the Quantum Fisher Information, providing strong numerical evidence for the 3D QKZM.

Together, these results establish NQS as a scalable framework for real-time dynamics in 3D quantum matter and as a reliable numerical counterpart to 3D quantum simulators.
\section{Numerical Method}
\subsection{Neural quantum states framework}
Given an initial state $|\Psi_0\rangle$, our goal is to compute the time-evolved state $|\Psi(t)\rangle$ by solving the Schrödinger equation $i\frac{d}{dt}|\Psi(t)\rangle = H(t) |\Psi(t)\rangle$, with $H(t)$ representing the Hamiltonian of a three-dimensional system on an $L\times L\times L$ lattice.
An exact numerical solution is intractable for the system sizes of interest due to the exponential growth of the Hilbert space dimension with the number of spins, $N=L^3$.

To tame the exponential scaling of the Hilbert space, we represent the quantum many-body state as a neural quantum state,
\(
|\Psi_{\boldsymbol{\theta}}\rangle=\sum_{s}\Psi_{\boldsymbol{\theta}}(s)\,|s\rangle,
\)
where the wave function \(\Psi_{\boldsymbol{\theta}}(s)\) is parametrized by an artificial neural network as
\(
\Psi_{\boldsymbol{\theta}}(s)=\exp\!\big[\eta_{\boldsymbol{\theta}}(s)\big],
\)
with \(\eta_{\boldsymbol{\theta}}(s)\) denoting the network output for configuration \(s\).
Here, the network weights \(\boldsymbol{\theta}\) serve as variational parameters.
A primary advantage of this approach is that, due to the universal approximation theorem~\cite{Hornik1991NeuralNetworks}, its accuracy can be systematically improved by increasing the size of the underlying neural network, enabling controlled convergence toward the exact many-body state.

Real-time evolution is performed using the time-dependent variational principle (TDVP), which projects the Schr\"odinger equation onto the variational manifold and yields an effective equation of motion
$\sum_k S_{jk}(\boldsymbol{\theta})\,\dot{\theta}_k = -\,i\,F_j(\boldsymbol{\theta}).$
Here $S$ is the covariance matrix of the variational derivatives $O_k(s)=\partial \ln \Psi_{\boldsymbol{\theta}}(s)/\partial \theta_k$ and $F$ is the connected correlator between $O_k$ and the local energy $E_{\textrm{loc}}(s)=\sum_{s'}\langle s|H|s'\rangle\Psi_{\boldsymbol{\theta}}(s')/\Psi_{\boldsymbol{\theta}}(s)$.
Expectation values of these quantities, as well as of the observables of interest, are estimated by averaging over configurations sampled from the probability distribution $|\Psi_{\boldsymbol{\theta}}(\boldsymbol{\sigma})|^2$ using the Markov-Chain Monte Carlo. 
We integrate the TDVP equations using a second-order Heun scheme with an adaptive time step~\cite{Schmitt2020PRL,SchmittHeyl2025arXiv}.

\subsection{3D ResNet-CNN architecture}
To represent the quantum state on a cubic lattice, our NQS employs a translationally equivariant convolutional neural network utilizing 3D kernels and residual connections.
The network maps an input spin configuration $S\equiv\{s_i\}$ on an $L\times L\times L$ lattice to the complex log-wavefunction $\ln\Psi_{\boldsymbol{\theta}}(S)$
[Fig.~\ref{fig:Figure0}(a)].
The network is built from a stack of $n$ residual blocks, with each residual block comprising two $3\times 3\times 3$ convolutional layers with unit stride and $2c$ output channels.
Each convolutional layer is followed by a Gaussian Error Linear Unit (GELU) activation, with a residual connection summing the block’s input and output.

We normalize each residual block by rescaling the residual contribution by a factor $1/\sqrt{i+1}$ (with $i$ the block index) to stabilize training for increasing depth.
Periodic boundary conditions are incorporated via circular padding in all convolutional layers to ensure translation equivariance on the 3D torus.
This ensures the network respects global translational invariance, making the architecture well-suited for investigating dynamics that preserve these symmetries.

The complex nature of the wavefunction is handled by a ``Pair Complex" layer that splits the final $2c$ output channels into real and imaginary parts, and pairs them to form $c$ complex channels.
These \(c\) channels are then exponentiated and summed to obtain a single complex scalar, after which we take the logarithm to produce \(\ln \Psi_{\boldsymbol{\theta}}(S)\).
For the numerical simulations presented herein, we fix $c=4$, and vary the network depth $n$ to control the network's expressivity.
Due to the weight-sharing property of the convolutional layers, the number of variational parameters remains independent of system size for a fixed depth and channel width, enabling simulations on large 3D lattices.

\section{Model}
We benchmark this numerical framework on the transverse-field Ising model (TFIM) defined on a cubic lattice of linear size $L$ with periodic boundary conditions. The Hamiltonian is given by
\begin{equation}
H = -J \sum_{\langle i j \rangle} \sigma_i^{z} \sigma_j^{z}
      - h \sum_i \sigma_i^{x},
\label{eq:Hamiltonian}
\end{equation}
where $\sigma_i^{\alpha}$ ($\alpha = x,z$) are Pauli matrices acting on site $i$, the first sum runs over nearest-neighbor pairs $\langle i j \rangle$, $J>0$ denotes the ferromagnetic nearest-neighbor coupling, and $h$ is the transverse field.
The model exhibits a quantum phase transition at a critical field $h_c=5.158136\, J$ separating ferromagnetic and paramagnetic phases~\cite{Bloete2002PRE}.

Using the NQS framework described in the previous section, we investigate the dynamics of this model in two scenarios: sudden quenches, where $H$ is time-independent for $t>0$, and smooth ramps, where the Hamiltonian $H(t)$ varies continuously by tuning the parameters $J(t)$ and $h(t)$, allowing us to probe the 3D quantum Kibble--Zurek mechanism.

Having introduced the NQS framework, the underlying network architecture, and the physical model, we now present the results of our NQS simulations.

\section{Quench dynamics}
\subsection{Collapse-and-revival dynamics}
Starting from a symmetry-broken ferromagnetic product state $|\Psi_0\rangle = \bigotimes_{\ell}|\uparrow\rangle_{\ell}$, we perform a global quench deep into the paramagnetic regime at $h=2h_c$.
To resolve the numerical challenges associated with the initial state as a delta peak in the computational $\sigma^z$ basis, we first rotate the Hamiltonian about the $\sigma^y$ axis by $\pi/2$ such that the initial state has uniform amplitudes.  
To further extend the reach of our simulations to longer timescales, we evolve the NQS in a rotating frame. A similar strategy was also explored in Ref.~\cite{Mendes2024PRX}.
Specifically, we decompose the Hamiltonian as $H=H_0+V$ with the transverse-field part
$H_0=-h\sum_i\sigma_i^x$ and interaction
$V=-J\sum_{\langle ij\rangle}\sigma_i^z\sigma_j^z$,
and move to the interaction picture with $U(t)=e^{iH_0 t}$.
This incorporates the global Larmor precession generated by $H_0$ directly at the level of the basis, so that the remaining nontrivial dynamics are primarily governed by the interactions in $V$.

Utilizing this framework, we obtain converged NQS dynamics using the 3D ResNet--CNN with depth $n=4$, in agreement with the $n=3$ results over the timescales presented. 
We show this convergence in Appendix~\ref{app:convergence:cr}.
The resulting dynamics for a $5\times5\times5$ system are shown in Fig.~\ref{fig:Figure0}(B).
We observe pronounced collapse-and-revival oscillations in the longitudinal magnetization $\langle\sigma^z_i(t)\rangle$, with an amplitude that gradually decays in time as interactions drive the system toward relaxation.
Remarkably, the NQS also reliably captures the buildup of strong multipartite entanglement, quantified by the quantum Fisher information (QFI) density associated with the longitudinal spin operator $\hat{\mathcal O}_z=\sum_i \sigma_i^z,$
\begin{equation}
f_Q[\hat{\mathcal O}_z]
=
\frac{1}{N}
\left(
\langle \hat{\mathcal O}_z^2\rangle-\langle \hat{\mathcal O}_z\rangle^2
\right),
\label{eq:QFI}
\end{equation}
where $N$ is the total number of spins.
We find that $f_Q$ increases with each collapse-and-revival oscillation, reaching values as large as $29$.
This implies multipartite entanglement extending over at least $30$ spins~\cite{Luca2018RMP}, consistent with the enhanced connectivity of the cubic lattice, which accelerates the spread of quantum correlations and entanglement in 3D.
These results demonstrate that the NQS framework can accurately capture both coherent quantum dynamics and strong multipartite entanglement growth in 3D.

\subsection{A sudden quench to the quantum critical point}
We now benchmark NQS in one of the most demanding regimes for real-time simulations, dynamics following a sudden quench to criticality.
We prepare the system in the paramagnetic product state
$|\Psi_0\rangle=\bigotimes_{\ell}|\rightarrow\rangle_{\ell}$
and abruptly quench the transverse field to the critical value $h_c$.
We find systematic convergence of the NQS dynamics for network depths $n=3$ and $n=4$.
We show this convergence in Appendix~\ref{app:convergence:qcp}.
In Fig.~\ref{fig:Figure0}(C), we show the resulting dynamics for a $6\times6\times6$ lattice, focusing on the transverse magnetization $\langle\sigma^x_i(t)\rangle$ and the quantum Fisher information density $f_Q$ associated with the longitudinal spin operator $\hat{\mathcal{O}}_z = \sum_i\sigma^z_i$ as defined in Eq.~\ref{eq:QFI}.
Notably, the NQS captures the strong growth of $f_Q$, signaling the buildup of nontrivial multipartite entanglement during the quench.
By accessing such large system sizes and long evolution times, our results establish new state-of-the-art benchmarks for quench dynamics in 3D.
\begin{figure*}[t]
\centering
\includegraphics[width=1.0\textwidth]{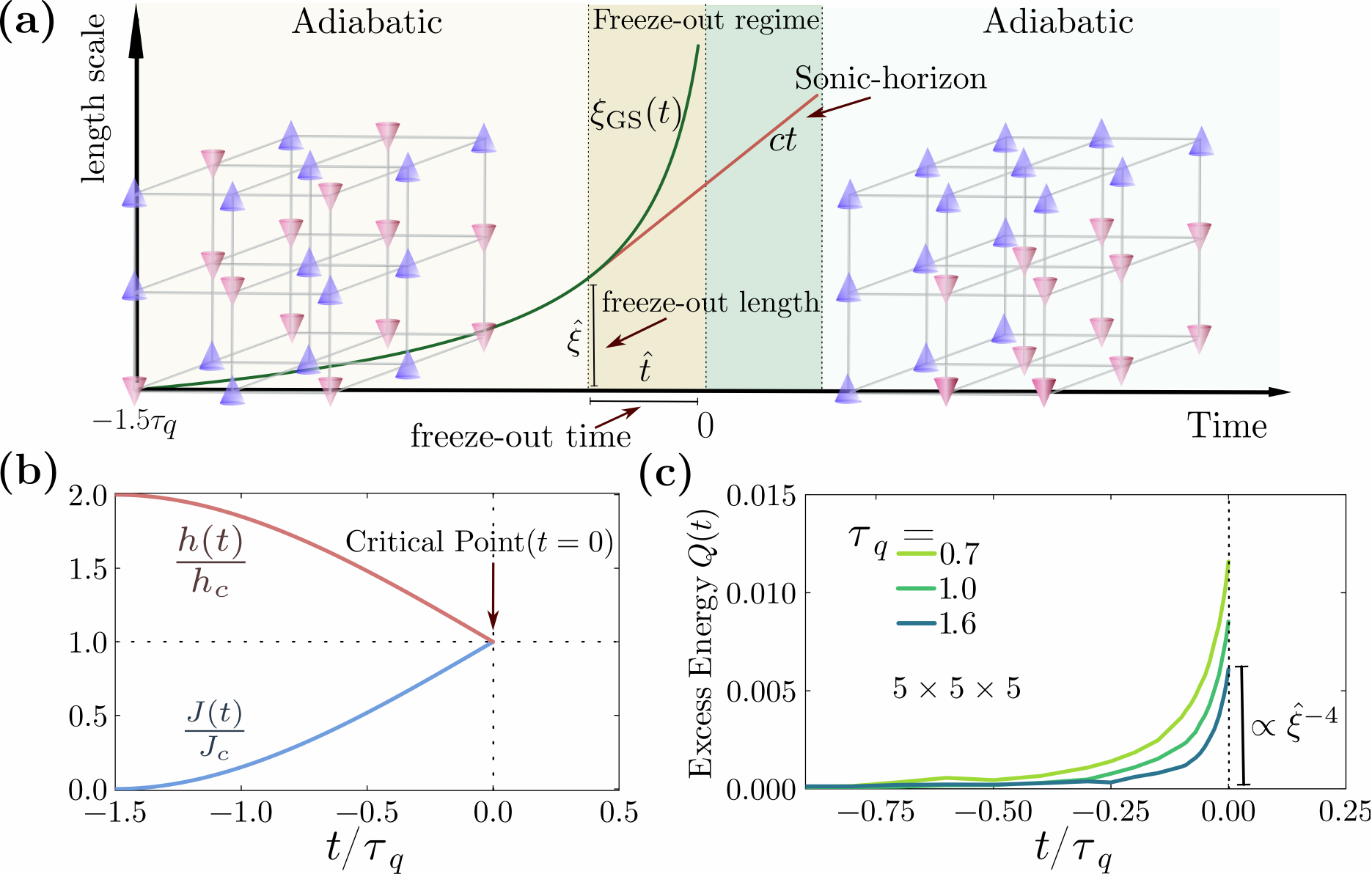}
\caption{\textbf{QKZM schematic and ramp protocol.} 
\textbf{(a)} The instantaneous ground-state correlation length $\xi_{\text{GS}}(t)$ (green) is compared against the causal bound $ct$ (red). When the growth rate of the dynamical correlation length reaches the maximal speed $c$, the system can no longer follow the instantaneous ground state and crosses over into the freeze-out regime, defining the freeze-out time $\hat{t}$ and the freeze-out length $\hat{\xi}$.
\textbf{(b)} The smooth ramp protocol: the transverse field $h(t)/h_c$ and Ising coupling $J(t)/J_c$ are varied to reach the critical point at $t=0$. The ramp time $\tau_q$ sets the ramp speed, which determines the growth rate of the dynamical correlation length.
\textbf{(c)} Excess energy $Q(t)$ for various $\tau_q$. At the critical point ($t=0$), the excess energy follows the universal scaling $Q(0) \propto \hat{\xi}^{-4}$, characteristic of the upper critical dimension $d+z=4$.}
\label{fig:Figure1}
\end{figure*}
\section{Quantum Kibble–Zurek Mechanism}
Having established the capability of NQS to simulate 3D real-time quantum dynamics following sudden quenches, we now turn to finite-rate quenches, in which a parameter of the Hamiltonian is varied continuously in time.
Such ramps across a quantum critical point (QCP) provide a natural setting to probe universal nonequilibrium dynamics governed by the critical properties of the phase transition.
The resulting dynamics is captured by the quantum Kibble--Zurek mechanism (QKZM), which describes the universal breakdown of adiabatic evolution in the vicinity of a quantum critical point.

\subsection{Physical picture}
Originally introduced in the context of topological-defect formation in the early universe~\cite{Kibble1976JPhysA}, the Kibble--Zurek framework has since become a central paradigm for understanding universal nonequilibrium dynamics across a wide range of scenarios. 
These include condensed-matter systems exhibiting universal dynamics via thermal quenches across a classical phase transition~\cite{Zurek1985Nature}, as well as universal quantum dynamics driven by real-time quenches across a QCP~\cite{Zurek2005PRL,Dziarmaga2005PRL}.

In the quantum setting, one considers a system initially prepared in its ground state and then driven through a QCP by varying a control parameter $\lambda$ in time.
At early times, the evolution remains adiabatic because the excitation gap is still large, consistent with the adiabatic theorem~\cite{Born1928ZPhys}.
As a result, the characteristic correlation length in the state follows the instantaneous ground-state correlation length $\xi_{\textrm{GS}}$.
As the critical point $\lambda_c$ is approached, however, $\xi_{\textrm{GS}}$ diverges.
At the same time, the spread of correlations in the real-time dynamics is constrained by a finite maximal propagation speed $c$.
Consequently, the system can no longer adiabatically track the instantaneous ground state.
This occurs at a characteristic time scale $\hat{t}$, commonly referred to as the `freeze-out' time, when the growth rate of the ground-state correlation length matches the maximal propagation speed, $|\partial\xi_\textrm{GS}/\partial t| \propto c$, thereby defining a sonic horizon beyond which dynamics become nonadiabatic.
This sonic-horizon picture~\cite{Sadhukhan2020PRB} of the Kibble--Zurek mechanism is illustrated in Fig.~\ref{fig:Figure1}(a).

\subsection{Scaling framework}
The connection to universality follows from the critical behavior of both the maximal propagation speed \(c\) and the ground-state correlation length \(\xi_{\textrm{GS}}\) near the QCP.
For correlated regions of linear size \(\xi\), the relevant propagation velocity obeys the scaling relation \(v \propto \xi^{-(z-1)}\), where \(z\) is the dynamical critical exponent.
Hence, for \(z=1\), as in quantum Ising models, the maximal propagation speed remains finite and approaches a constant near the QCP.
The ground-state correlation length exhibits a universal dependence on the dimensionless distance from criticality,
\(r(t)=\left(\lambda(t)-\lambda_c\right)/{\lambda_c},\)
typically through the scaling relation
\(\xi_{\textrm{GS}} \propto r^{-\nu},\)
where \(\nu\) is the universal correlation-length exponent.
In special cases, such as at the upper critical dimension, this leading power-law behavior is supplemented by logarithmic factors.

Given these equilibrium properties, the freeze-out time $\hat{t}$ is determined as a function of the quench time $\tau_q$ from the sonic-horizon condition, $\xi_{\textrm{GS}}(\hat{t}) \sim c\,\hat{t}$.
The associated freeze-out length is then $\hat{\xi} = \xi_{\textrm{GS}}(\hat{t})$, which, using dynamic scaling, is equivalently $\hat{\xi} \sim \hat{t}^{1/z}.$
These scales govern the universal nonequilibrium scaling of quantities such as equal-time correlation functions and the excess energy, providing the framework for the numerical identification of Kibble–Zurek scaling.

\subsection{Scaling framework for the 3D TFIM}
\subsubsection*{Ramp protocol}
We now apply this framework to the 3D transverse-field Ising model (TFIM) described by Eq.~\eqref{eq:Hamiltonian}.
We first prepare the system in the paramagnetic ground state at \(h=2h_c\) and \(J=0\).
We then vary both the transverse field \(h(t)\) and the Ising coupling \(J(t)\) according to
\begin{align*}
    \frac{J(t)}{J_c} = \left(1 + \frac{t}{\tau_q} - \frac{4}{27}\frac{t^3}{\tau_q^3}\right),\,\,\,\,
    \frac{h(t)}{h_c} = \left(1 - \frac{t}{\tau_q} + \frac{4}{27}\frac{t^3}{\tau_q^3}\right).
\end{align*}
This quench begins at \(t=-3\tau_q/2\), and crosses the critical point at \(t=0\), where \(J(0)=J_c=1\) and \(h(0)=h_c\), as illustrated in Fig.~\ref{fig:Figure1}(B).
The quench time \(\tau_q\) sets the speed of the quench with larger \(\tau_q\) corresponding to a slower quench.
We employ this nonlinear ramp in order to suppress excitations generated at the beginning of the evolution, while near $t=0$, this ramp effectively becomes a linear quench $r(t)\sim t/\tau_q$.

To characterize the universal behavior of these dynamics, we next determine the freeze-out time $\hat{t}$ as a function of the quench time $\tau_q$.
\subsubsection*{Refined freeze-out scales}
This requires understanding the critical scaling of both the maximal propagation speed and the ground-state correlation length.
Since the dynamical critical exponent is $z=1$, the maximal propagation speed approaches a constant near criticality.
The nontrivial part of the scaling analysis, therefore, enters through the ground-state correlation length, whose behavior is captured by the scalar \(\phi^4\) field theory in effective dimensions $d+z=4$, placing the system exactly at its upper critical dimension.
Below the upper critical dimension, critical behavior is governed by nontrivial universal exponents, whereas above it, the leading critical behavior is described by mean-field exponents.
Exactly at the upper critical dimension, the asymptotic scaling of the correlation length exhibits multiplicative logarithmic factors to the leading mean-field power law~\cite{Cardy1996, Justin2002}.

More precisely, this asymptotic scaling of the correlation length is obtained from a renormalization-group (RG) analysis of the field theory.
While the exact RG flow is generally not analytically tractable, a leading-order one-loop treatment from perturbative RG already determines the asymptotic form of \(\xi_{\mathrm{GS}}(r)\).
However, this asymptotic form often does not provide a sufficient quantitative description of numerically or experimentally accessible regimes.
This is particularly important at the upper critical dimension, where the convergence to the asymptotic scaling form is exceptionally slow: sub-leading corrections vanish only as inverse powers of \(\ln(1/r)\), rather than through the typically algebraic corrections, \(\mathcal{O}(r^{\,w})\)~\cite{Goldenfeld1992}, encountered below the upper critical dimension.
We therefore go beyond the leading asymptotic analysis and derive a more accurate form of \(\xi_{\mathrm{GS}}(r)\) by integrating the RG flow equations to two-loop order~\cite{Justin2002,kleinert1991PLB}.
See Appendix.~\ref{app:RG} for the details of this RG analysis.

We then obtain the freeze-out time $\hat{t}$ by substituting \(r \to \hat{t}/\tau_q\) into the expression for the ground-state correlation length and imposing the sonic-horizon condition \(\xi_{\mathrm{GS}}(\hat{t}) \sim c\,\hat{t}\).
The resulting scaling form is
\begin{align}
    \hat{t} \propto \tau_q^{1/3} [\ln (C\tau_q)]^{1/9}
    \left( 1 + \frac{\mathcal{G}_\mu(\ln C\tau_q)}{\ln (C\tau_q)} \right),
    \label{eq:refined_scaling}
\end{align}
where the function \(\mathcal{G}_\mu(x) = \frac{5}{27} + \frac{1}{3\mu} - \frac{34}{81}\ln\left(1 + \frac{\mu}{3}x\right)
\) captures sub-leading corrections.
This expression shows that the mean-field power law, \(\hat{t} \sim \tau_q^{1/3}\), is modified by an additional \((\ln \tau_q)^{1/9}\) factor, together with sub-leading corrections that decay as inverse powers of the logarithm.
Once \(\hat{t}\) is determined, the freeze-out length follows as \(\hat{\xi} \sim \hat{t}^{1/z}\).

In Eq.~\eqref{eq:refined_scaling}, \(C\) and \(\mu\) are nonuniversal constants set by the microscopic details of the model and therefore cannot be determined within the field-theoretical approach.
We thus treat them as fitting parameters in the numerical data collapse.
Importantly, these parameters are independent of the observables under consideration.
Once these constants are fixed, for instance, by collapsing the data for the correlation length, they must consistently describe the scaling of other independent quantities, such as the excess energy. 
This provides a rigorous validation of our scaling framework.

In the next section, we show how this scaling form is used to perform data collapses of different quantities across multiple system sizes.

\subsection{Kibble--Zurek scaling forms for observables}
Under the Kibble--Zurek framework, a relevant operator \(\hat O\) is expected to obey the universal scaling form~\cite{Anushya2012PRB, Anna2016PRB, Dziarmaga2022PRL}
\begin{equation}
    \langle \hat O(t,R;\tau_q)\rangle
    =
    \hat{\xi}^{-\Delta_O}\,
    \mathcal{F}_O\!\left(
    \frac{t}{\hat t},
    \frac{R}{\hat\xi}
    \right),
\end{equation}
where \(\Delta_O\) is the scaling dimension of \(\hat O\).
Here, \(t/\hat t\) and \(R/\hat\xi\) are the natural scaling variables associated with time and distance, respectively.
At fixed scaled time and scaled distance, this implies \(\langle \hat O\rangle \sim \hat\xi^{-\Delta_O}\).
The dependence on the quench time \(\tau_q\) therefore enters only through the emergent freeze-out scales \(\hat t\) and \(\hat\xi\).

For spatially resolved observables such as correlation functions, this form directly implies data collapse when the rescaled observable is plotted as a function of \(R/\hat\xi\), and this collapse can be tested across different system sizes.
For spatially integrated quantities such as the excess energy, however, a finite-size scaling analysis is required to establish collapse across system sizes.
At the upper critical dimension, finite-size scaling~\cite{kenna2004NPB} is modified by multiplicative logarithmic corrections.
In particular, the finite-size correlation length scales as \(\xi_L \sim L (\ln L)^{1/4}\), so that finite-size effects enter through the logarithmically dressed system size \(L_{\mathrm{eff}} \sim L (\ln L)^{1/4}\).

Accordingly, finite-size data collapse is tested by plotting \(L_{\mathrm{eff}}^{\Delta_O} O\) as a function of \(\hat\xi/L_{\mathrm{eff}}\).
If the scaling hypothesis holds, data for different system sizes collapse onto a single universal curve, with the remaining dependence fully encoded in the universal scaling function.

Next, we show that our NQS simulations for the 3D TFIM exhibit a compelling data collapse consistent with these scaling forms. 
\subsection{Numerical results}
\begin{figure}[t]
\centering
\includegraphics[width=0.48\textwidth]{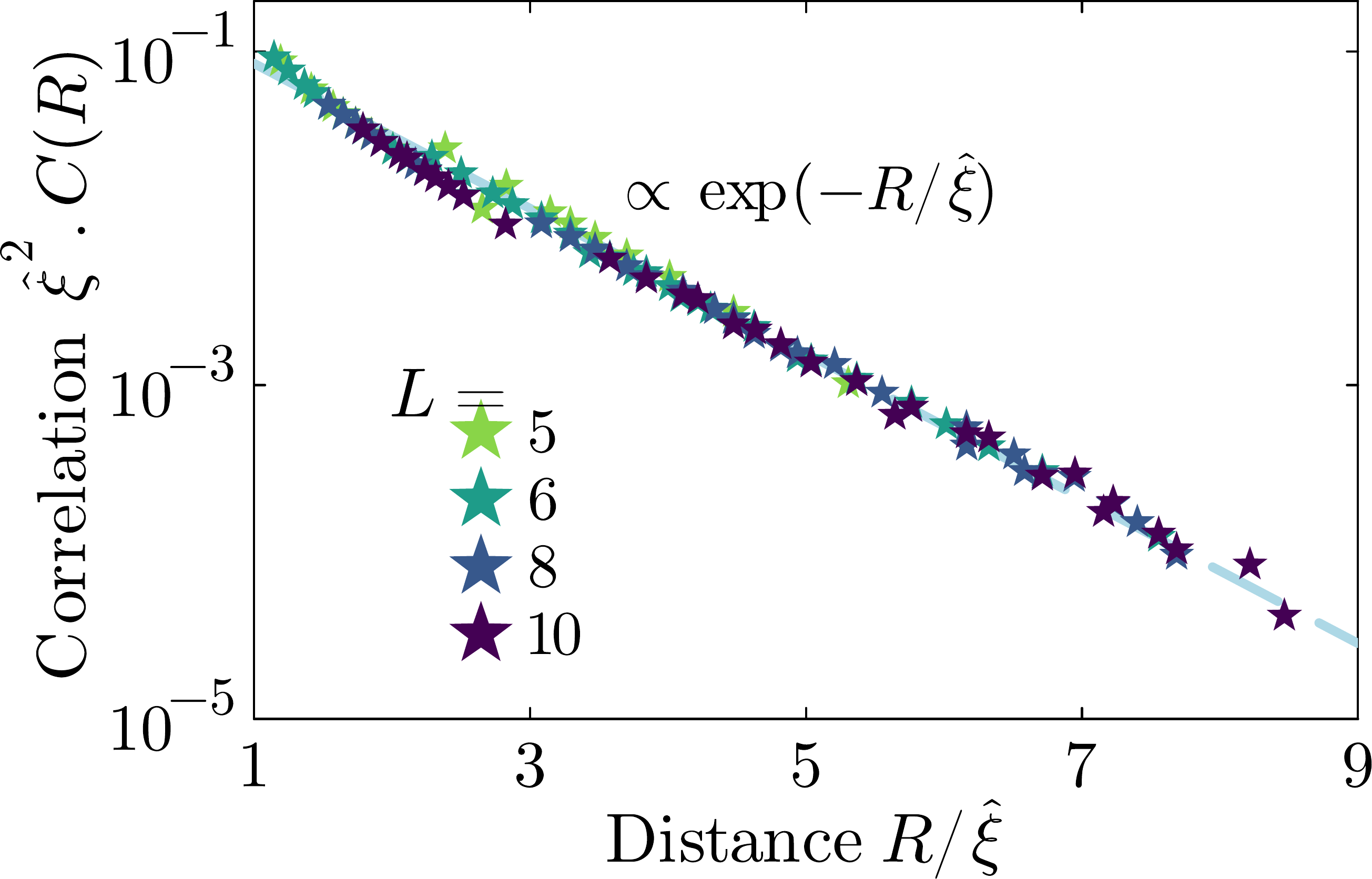}
\caption{
Correlation-function collapse for different system sizes and different ramp times.
The rescaled correlation function $\hat{\xi}^{2} C(R)$, evaluated at $t=0$, is plotted as a function of the scaled distance $R/\hat{\xi}$.
The data collapse onto a single curve and exhibit the expected exponential form, $C(R)\propto e^{-R/\hat{\xi}}$(light blue line).
}
\label{fig:figure2}
\end{figure}

\subsubsection*{Scaling collapse of the correlation function}
We first analyze the equal-time correlation function
\begin{equation*}
    C(t, R) = \langle\psi(t)|\sigma^z_0\sigma^z_R|\psi(t)\rangle,
\end{equation*}
evaluated at the critical point, $t=0$.
Because the finite-rate sweep drives the system out of adiabaticity, the state at $t=0$ is not itself critical.
Accordingly, the correlation function remains short-ranged and decays exponentially with distance.

To test whether the freeze-out length $\hat{\xi}$ in Eq.~\eqref{eq:refined_scaling} correctly captures the non-equilibrium correlation length, we perform a data collapse of the correlation data. 
Specifically, we plot $\hat{\xi}^{\Delta} C(R)$ as a function of the rescaled distance $R/\hat{\xi}$ for various ramp times and system sizes. 
Following the scaling dimension of the two-point correlator $\langle \hat{\sigma}_i^z \hat{\sigma}_j^z \rangle$, we set $\Delta = d+z - 2 + \eta = 2$ given that $d=3, z=1,$ and $\eta=0$ in the 3D quantum case~\cite{sachdev1999quantum}.
As shown in Fig.~\ref{fig:figure2}, the data exhibit a compelling collapse onto the universal scaling form 
\begin{equation*}
    \hat{\xi}^{\Delta} C(R) = \mathcal{F}(R/\hat{\xi}) \propto e^{-R/\hat{\xi}}.
\end{equation*}
To obtain this collapse, we perform a numerical fit, described in Appendix~\ref{app:collapse}, which determines the nonuniversal constants $C=81.189$ and $\mu=0.194$ entering the freeze-out length $\hat{\xi}$.
To demonstrate the robustness of the scaling framework, we keep these constants fixed for all other observables.
\begin{figure}[t]
\centering
\includegraphics[width=0.48\textwidth]{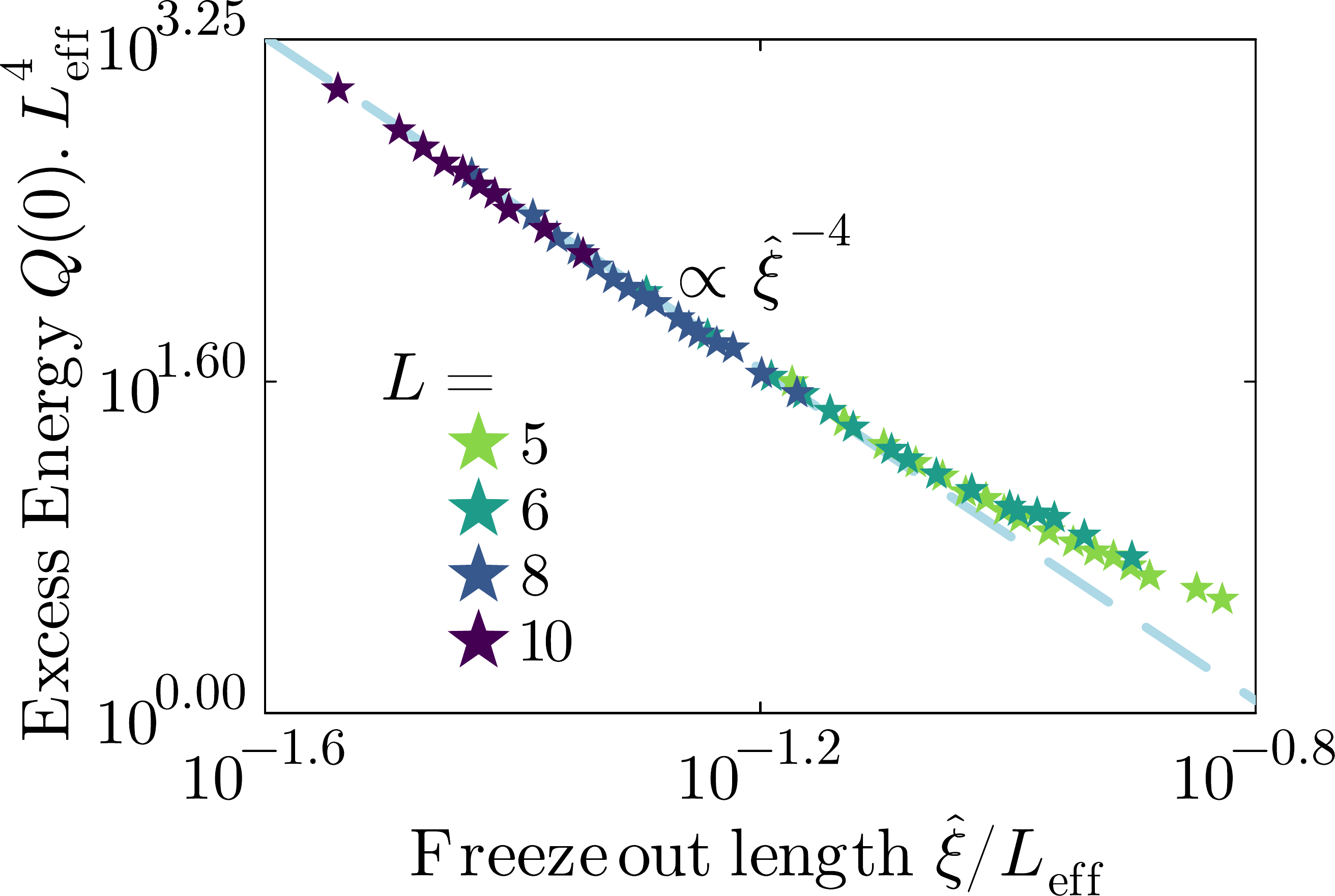}
\caption{
Excess-energy collapse for different system sizes.
Plotting \(Q L_{\mathrm{eff}}^{4}\) at \(t=0\) versus \(\hat{\xi}/L_{\mathrm{eff}}\) yields a finite-size scaling collapse consistent with the asymptotic Kibble--Zurek prediction \(Q \sim \hat{\xi}^{-4}\) (light blue line).
}
\label{fig:figure3}
\end{figure}
\begin{figure*}[t]
\centering
\includegraphics[width=\textwidth]{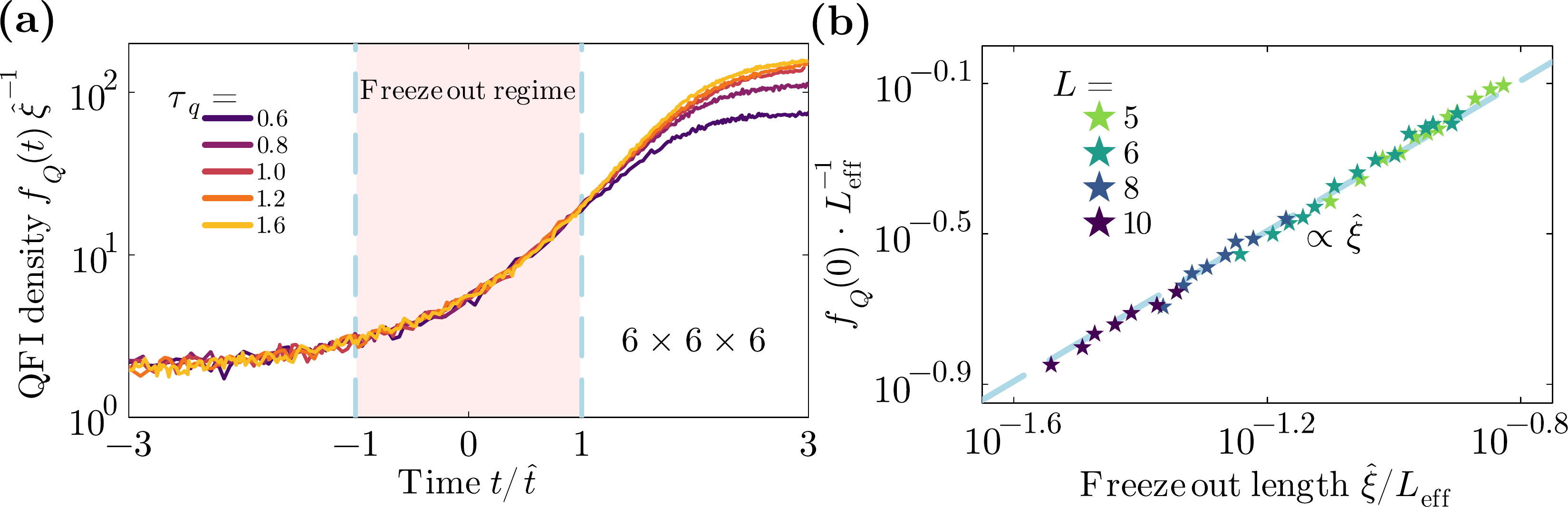}
\caption{
Universal dynamics of multipartite entanglement.
(a) Time evolution of the rescaled QFI density $f_Q(t)/\hat{\xi}$ for different ramp times in a $6\times 6\times 6$ system, plotted versus the rescaled time $t/\hat{t}$.
The shaded region marks the Kibble--Zurek regime, where the data collapse onto a universal curve.
(b) Plotting $f_Q(0)\cdot L^{-1}_\mathrm{eff}$ as a function of the scaled freeze-out length $\hat{\xi}/L_\mathrm{eff}$ yields a data collapse consistent with the expected scaling $f_Q(0)\sim \hat{\xi}$ (light blue line).
}
\label{fig:figure4}
\end{figure*}
\subsubsection*{Scaling collapse of the excess energy}
We next analyze the excess energy density $Q$, defined as the difference between the instantaneous energy density and the corresponding ground-state energy density:
\begin{equation*}
Q(t) = \frac{1}{L^3}\left(\langle\psi(t)|H(t)|\psi(t)\rangle - \langle\psi_{\mathrm{GS}}|H(t)|\psi_{\mathrm{GS}}\rangle\right).
\end{equation*}
The scaling dimension of $Q$ is $\Delta_Q = d+z = 4$~\cite{DeGrandi2010book}.
In the scaling analysis, we use the same values of the nonuniversal parameters $C$ and $\mu$ as determined from the correlation-function collapse.
Fig.~\ref{fig:figure3} shows the rescaled excess energy $Q\,L_{\mathrm{eff}}^{4}$ at $t=0$ as a function of the scaled freeze-out length $\hat{\xi}/L_{\mathrm{eff}}$.
We find that the data for all system sizes collapse onto a single universal curve, demonstrating the universal scaling of the excess energy.

In the regime $\hat{\xi}/L_\mathrm{eff}\ll1$, we observe compelling evidence for the expected power-law scaling $Q \sim \hat{\xi}^{-4}$, in agreement with the scaling dimension $\Delta_Q = 4$.
For larger values of \(\hat{\xi}/L_\mathrm{eff}\), however, systematic deviations from Kibble--Zurek scaling emerge, signaling a crossover out of the Kibble--Zurek regime.
A more detailed discussion of the large-\(\hat{\xi}/L_\mathrm{eff}\) behavior is given in Appendix~\ref{app:numerics:qkzm}.

\subsubsection*{Universal dynamics of multipartite entanglement}
Having established universal scaling in the correlation function and excess energy, we finally ask how Kibble--Zurek scaling is encoded in the entanglement structure of the quantum many-body state.
To address this question, we analyze the quantum Fisher information (QFI) density associated with the longitudinal spin operator $\hat{\mathcal O}_z=\sum_i \sigma_i^z$ as defined in Eq.~\ref{eq:QFI}, which serves as a probe of multipartite entanglement.
Its scaling dimension is $\Delta_{f_Q}=2\Delta_{\hat{\sigma}_z}-d=-1$, where $\Delta_{\hat{\sigma}_z}=\frac{d+z - 2 + \eta}{2} = 1$, given that $d=3$, $z=1$ and $\eta=0$ in the 3D quantum case~\cite{hauke2016natphys}.
Figure~\ref{fig:figure4}(a) shows the real-time evolution of the rescaled QFI density, $f_Q(t)/\hat{\xi}$, as a function of the rescaled time $t/\hat{t}$ for several ramp times.
As the system is driven through the critical region, the QFI grows strongly, signaling the buildup of multipartite entanglement over increasingly large blocks of spins.
Under this rescaling, the data collapse throughout the Kibble--Zurek regime, $-\hat{t}<t<\hat{t}$, consistent with the scaling form
\[
f_Q(t,\tau_q)=\hat{\xi}\,\mathcal{F}_{f_Q}(t/\hat{t}).
\]
The deviations observed for $t>\hat{t}$ reflect the crossover out of the Kibble--Zurek regime and the onset of late-time coarsening dynamics.

To further quantify this behavior, Fig.~\ref{fig:figure4}(b) shows the QFI density at the critical point, $f_Q(0)$, rescaled by the linear system size as $f_Q(0)/L_{\mathrm{eff}}$, and plotted as a function of the scaled freeze-out length $\hat{\xi}/L_{\mathrm{eff}}$.
The data for all system sizes exhibit a compelling collapse following the scaling relation $f_Q(0)\sim \hat{\xi}$.
This demonstrates that the universal dynamics of multipartite entanglement is governed by the same freeze-out scale that controls the correlation function and excess energy.
Compared to the excess energy, the QFI density follows the Kibble--Zurek scaling form over a broader range of \(\hat{\xi}/L_{\mathrm{eff}}\).
These results further establish NQS as a robust framework for accessing universal entanglement dynamics in 3D quantum many-body systems.

\section{Conclusions and outlook}
In this work, we show that neural quantum states offer an efficient approach to simulating real-time dynamics
in 3D quantum many-body systems”.
By extending CNN-based NQS architectures to 3D lattices, we performed large-scale simulations of the transverse-field Ising model for system sizes of up to $10^3$ quantum spins.

Using this approach, we investigated two complementary classes of nonequilibrium protocols: sudden quenches and finite-rate quenches across the quantum critical point.
For sudden quenches, our results demonstrate that NQS can accurately capture coherent real-time dynamics in 3D, including nontrivial many-body signatures at system sizes far beyond the reach of conventional exact methods.
For finite-rate quenches, we showed that the same framework can resolve universal nonequilibrium dynamics governed by the quantum Kibble--Zurek mechanism.

Beyond the methodological advance, the main conceptual contribution of this work is the first numerical demonstration of the quantum Kibble–Zurek mechanism in the three-dimensional
transverse-field Ising model at the upper critical dimension.” is the first numerical demonstration of the quantum Kibble--Zurek mechanism in the 3D transverse-field Ising model, which lies at the upper critical dimension.
In this regime, the universal scaling structure is modified by logarithmic corrections, making the quantitative analysis significantly more subtle than in lower-dimensional settings.
To address this, we derived refined scaling laws from the renormalization-group flow equations to include the sub-leading inverse-logarithmic terms along with the asymptotic scaling behavior.
The resulting theoretical predictions are in excellent agreement with our NQS simulations across multiple observables, including the correlation function, excess energy, and quantum Fisher information.
This combined analytical and numerical analysis allowed us to establish the quantum Kibble--Zurek mechanism in 3D.

The present work opens several natural directions for future research.
The framework developed here can be applied to a broad range of open problems, including dynamical critical phenomena in other 3D models, such as 3D Rydberg atoms~\cite{barredo2018Nature}, and nonequilibrium dynamics beyond the Kibble--Zurek paradigm such as coarsening dynamics~\cite{manovitz2025nature}.
Another particularly promising direction is the quantitative study of universal entanglement dynamics in regimes directly accessible to programmable quantum simulators.
Moreover, our results reported here for 3D quenches, universal scaling, and multipartite entanglement furnish concrete reference results for validating both quantum simulators and other classical approaches to quantum many-body dynamics.

\section*{Data Availability}
The data to generate all figures in this article will be made available via Zenodo~\cite{dataset_3DQKZM}.

\section*{Acknowledgments}
This project has received funding from the European Research Council (ERC) under the European Union’s Horizon 2020 research and innovation programme (Grant Agreement No.~853443). This work was supported by the German Research Foundation (DFG) via project 492547816 (TRR~360) and via project 530080784. We gratefully acknowledge the scientific support and high-performance computing resources provided by the LiCCA HPC cluster at the University of Augsburg (co-funded by the DFG, Project ID 499211671), the Erlangen National High Performance Computing Center (NHR@FAU, Project No. nqsQuMat), and the GCS Supercomputer JUWELS at the Jülich Supercomputing Centre. The LiCCA cluster is co-funded by the DFG (Project ID 499211671). NHR funding is provided by federal and Bavarian state authorities, and the NHR@FAU hardware is partially funded by the DFG (Grant No.~440719683).

\appendix

\section{Numerical details and convergence benchmarks}
\label{app:numerics}
In this Appendix, we summarize the numerical implementation and present convergence benchmarks for the real-time dynamics discussed in the main text.

The simulations are performed using the \texttt{jVMC} package~\cite{Schmitt2022} with GPU acceleration.
Throughout this work, we mainly consider network depths \(n=2\), \(n=3\), and \(n=4\), corresponding to \(5424\), \(8896\), and \(12368\) network parameters, respectively.

\subsection{Collapse-and-revival dynamics}
\label{app:convergence:cr}

\begin{figure*}[!t]
\centering
\includegraphics[width=\textwidth]{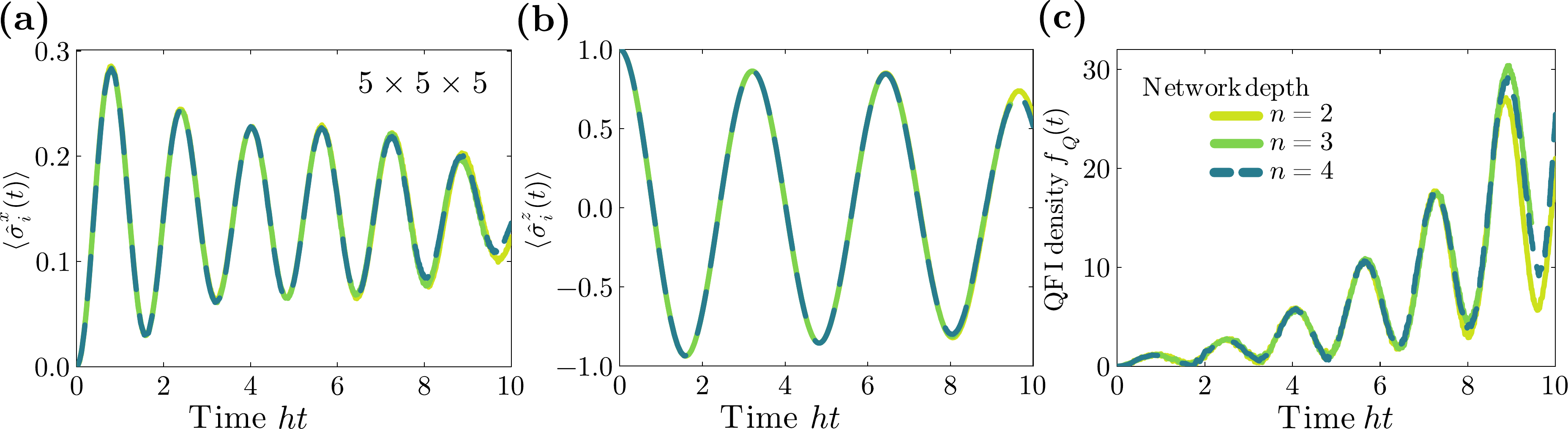}
\caption{Convergence of NQS for collapse-and-revival dynamics.
The panels illustrate the numerical convergence of the NQS dynamics on a \(5\times5\times5\) lattice with increasing network depth \(n\), for the expectation values of the following observables:
\textbf{(a)} transverse magnetization \(\langle \sigma_i^x(t)\rangle\),
\textbf{(b)} longitudinal magnetization \(\langle \sigma_i^z(t)\rangle\), and
\textbf{(c)} quantum Fisher information density \(f_Q(t)\).}
\label{fig:Figure6}
\end{figure*}

To assess the numerical reliability of the NQS approach in the collapse-and-revival regime, we study the convergence of the dynamics with increasing network depth \(n\).
Figure~\ref{fig:Figure6} shows results for a \(5\times5\times5\) lattice for three observables: the transverse magnetization \(\langle \sigma_i^x(t)\rangle\), the longitudinal magnetization \(\langle \sigma_i^z(t)\rangle\), and the quantum Fisher information density \(f_Q(t)\).
For these simulations, we use \(4\times10^4\) Monte Carlo samples at each time step.

We find that the NQS results converge systematically as the network depth is increased.
In particular, the collapse-and-revival oscillations in local observables, as well as the buildup of multipartite entanglement, are already captured consistently at moderate depth.
These results demonstrate that the chosen NQS architecture provides a controlled and reliable representation of the real-time dynamics in this regime.

\subsection{A sudden quench to the critical point}
\label{app:convergence:qcp}

\begin{figure*}[!t]
\centering
\includegraphics[width=\textwidth]{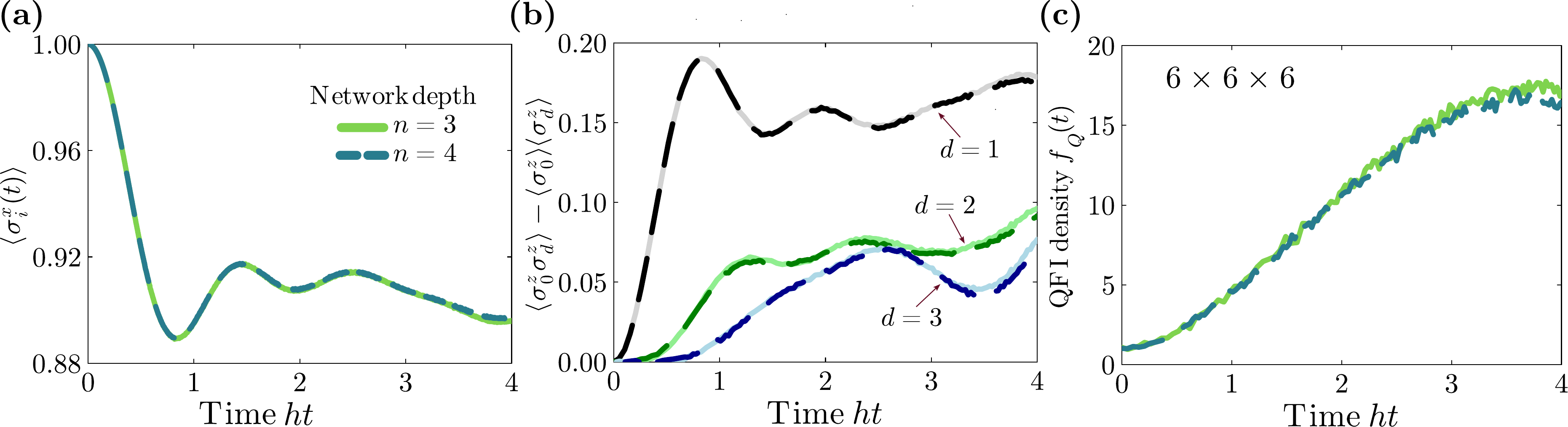}
\caption{Convergence of NQS for dynamics following a sudden quench to criticality.
The panels illustrate the numerical convergence of the NQS dynamics on a \(6\times6\times6\) lattice for network depths \(n=3\) and \(n=4\), for the expectation values of the following observables:
\textbf{(a)} transverse magnetization \(\langle \sigma_i^x(t)\rangle\),
\textbf{(b)} correlation function \(\langle
\sigma^z_0\sigma^z_d\rangle - \langle
\sigma^z_0\rangle\langle\sigma^z_d\rangle\), and
\textbf{(c)} quantum Fisher information density \(f_Q(t)\).}
\label{fig:Figure7}
\end{figure*}

We next examine the convergence of the NQS dynamics following a sudden quench to the quantum critical point.
Figure~\ref{fig:Figure7} compares results on a \(6\times6\times6\) lattice obtained with network depths \(n=3\) and \(n=4\).
We analyze the time evolution of the transverse magnetization \(\langle \sigma_i^x(t)\rangle\), the correlation function \(\langle
\sigma^z_0\sigma^z_d\rangle - \langle
\sigma^z_0\rangle\langle\sigma^z_d\rangle\), and the quantum Fisher information density \(f_Q(t)\).
For these simulations, we use \(10^5\) Monte Carlo samples at each time step.

Across all observables, the results for the two network depths agree very well over the time window shown, indicating that the NQS dynamics is well converged with respect to network expressivity.
This demonstrates that our NQS framework remains reliable even in this challenging dynamical regime.

\subsection{Finite-rate quenches -- QKZM}
\label{app:numerics:qkzm}
\begin{figure*}[t]
\centering
\includegraphics[width=\textwidth]{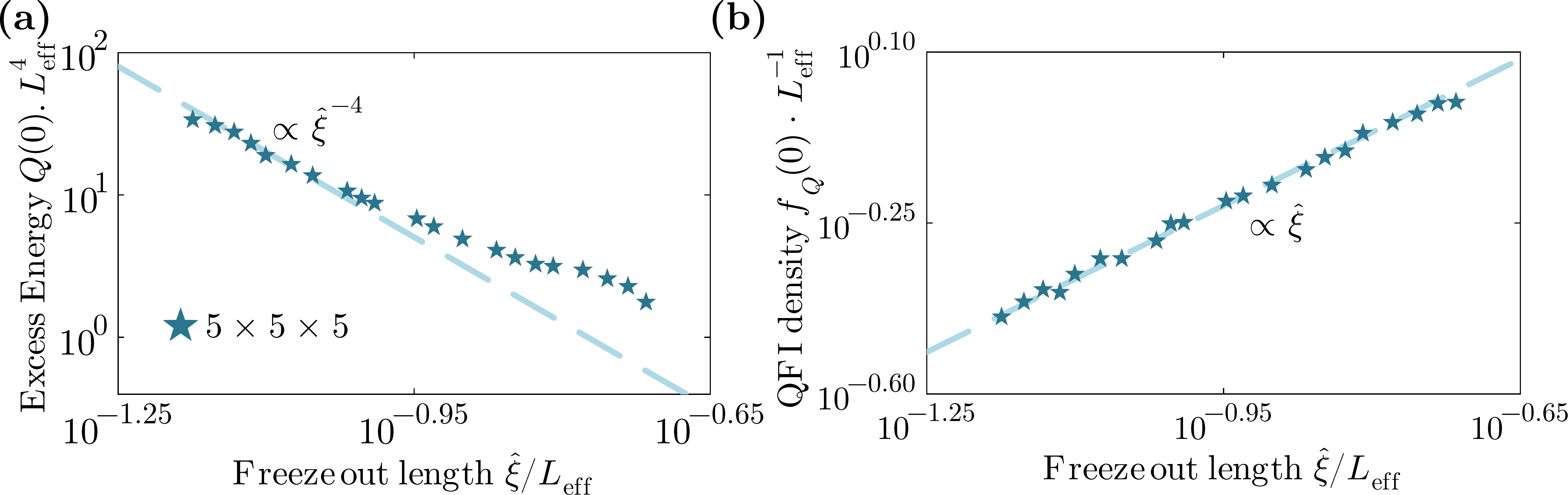}
\caption{Kibble--Zurek scaling with a linear ramp protocol for \(L=5\).
\textbf{(a)} Excess energy \(Q(0)\) and \textbf{(b)} quantum Fisher information density \(f_Q(0)\) as functions of the scaled freeze-out length \(\hat{\xi}/L_{\mathrm{eff}}\).
For \(\hat{\xi}/L_{\mathrm{eff}} \ll 1\), the data follow the expected Kibble--Zurek power-law scaling, as indicated by the dashed lines.
At larger values of \(\hat{\xi}/L_{\mathrm{eff}}\), systematic deviations for $Q$ signal a crossover out of the universal scaling regime toward finite-size-dominated dynamics, with possible onset of adiabatic behavior for the slowest ramps.}
\label{fig:figure8}
\end{figure*}
For the finite-rate quenches discussed in the main text, we consider system sizes from \(L=5\) to \(L=10\).
We use network depth \(n=3\) for most simulations, except for \(L=5\), where we use \(n=2\).
The time evolution is performed from \(t=-1.5\tau_q\) to \(t=0\), with \(t=0\) marking the point at which the ramp reaches criticality.
We consider quench times in the ranges
\(\tau_q\in[0.15,3.5]\) for \(L=5\),
\(\tau_q\in[0.15,4.0]\) for \(L=6\),
\(\tau_q\in[0.15,1.5]\) for \(L=8\), and
\(\tau_q\in[0.15,0.9]\) for \(L=10\).
The computational cost increases substantially with increasing \(\tau_q\), since slower ramps require longer real-time evolution and higher numerical accuracy.
Accordingly, we typically use between \(2\times10^4\) and \(10^5\) Monte Carlo samples at each time step, with larger sample numbers in the more demanding regime of slow ramps and large system sizes.

Evaluating the excess energy \(Q\) becomes increasingly challenging at large \(\tau_q\), since it involves subtracting two large quantities whose difference is very small.
The quantum Fisher information density \(f_Q\) poses a different challenge, as it involves summing many small long-distance correlation contributions and is therefore intrinsically noisy.
To obtain converged results for \(f_Q\), we use up to \(5\times10^5\) Monte Carlo samples.
Consequently, we obtain compelling numerical evidence for the expected Kibble--Zurek scaling behavior.

For sufficiently slow ramps, a finite system is expected to eventually crossover from the Kibble--Zurek regime to the adiabatic regime.
For the excess energy \(Q\), this crossover should be reflected in a stronger suppression than the asymptotic Kibble--Zurek form \(Q\sim \hat{\xi}^{-4}\).
However, Fig.~\ref{fig:figure8} shows that at intermediate values of \(\hat{\xi}/L_{\mathrm{eff}}\), \(Q\) instead begins to decay more weakly than predicted by Kibble--Zurek scaling.
A comparison with the quantum Fisher information density \(f_Q\) helps clarify this behavior.
While \(Q\) already deviates from its asymptotic scaling form, \(f_Q\) continues to follow the expected Kibble--Zurek behavior \(f_Q\sim \hat{\xi}\) over a broader range.
This suggests that the earlier deviation in \(Q\) is more naturally attributed to finite-size behavior, rather than the adiabatic crossover.

Since extending the smooth-ramp protocol to even larger \(\tau_q\) becomes increasingly expensive, we additionally consider for \(L=5\) a linear ramp protocol for longer quench times up to \(\tau_q=8.0\).
In this case, the dynamics starts at \(t=-\tau_q\) and reaches the critical point at \(t=0\), corresponding to a linear ramp in \(t/\tau_q\).
We consider this protocol because it is computationally less expensive, as the total evolution time is shorter.
The corresponding results for both \(Q\) and \(f_Q\) are shown in Fig.~\ref{fig:figure8}.

In the regime \(\hat{\xi}\ll L_{\mathrm{eff}}\), where Kibble--Zurek scaling is expected to hold, both observables recover the expected scaling behavior.
Upon increasing \(\hat{\xi}/L_{\mathrm{eff}}\), however, \(Q\) starts to deviate, while \(f_Q\) continues to follow the Kibble--Zurek form over a broader range.
Only for the slowest accessible ramps do we observe possible indications of the expected adiabatic crossover.
In this regime, the excess energy \(Q\) bends downward more strongly, while \(f_Q\) shows a tendency toward slower growth.
However, since only very few data points are available there, no firm conclusion can be drawn.
Overall, the data show that the expected Kibble--Zurek scaling is robustly recovered in the bulk regime \(\hat{\xi}\ll L_{\mathrm{eff}}\), while at larger \(\hat{\xi}/L_{\mathrm{eff}}\) the dynamics first becomes dominated by finite-size effects and only at the slowest ramps shows possible signatures of adiabatic crossover.

\section{KZ scaling from two-loop RG flow}
\label{app:RG}
In this Appendix section, we derive the explicit form of the freeze-out time $\hat{t}$ as a function of quench time $\tau_q$, focusing on the asymptotic scaling behavior and the inclusion of relevant sub-leading corrections.

The critical behavior of the 3D TFIM is described by the scalar \(\phi^4\) field theory in $d+z=4$ dimensions. The action corresponding to this theory is:
\begin{align*}
S = \int d^4x \left[
\frac{1}{2} (\nabla \phi)^2
+ \frac{r}{2}\phi^2
+ \frac{16\pi^2}{4!}u \, \phi^4
\right],
\end{align*}
where \(r\) represents the distance from criticality and \(u\) is the quartic interaction. 
We analyze the evolution of these couplings with respect to the logarithmic scale \(\ell\). 
Specifically, we employ the RG flow equations up to two-loop order~\cite{kleinert1991PLB} for the quartic coupling $u$ and quadratic coupling $r$:
\begin{align*}
\frac{du}{d\ell} = -\frac{3}{2}u^2 + \frac{17}{6}u^3, \ \ \ \
\frac{dr}{d\ell} = \left[ 2 - \frac{1}{2}u + \frac{5}{12}u^2 \right] r.
\end{align*}
Here, the fixed point is the Gaussian fixed point at \(u^*=0\), and substituting this into the flow equation for \(r\) recovers the leading mean-field result \(\nu=1/2\).
However, this is not sufficient, because the quartic coupling \(u\) approaches zero only logarithmically under RG flow, making it marginally irrelevant.
Thus, this generates multiplicative logarithmic corrections, as well as sub-leading inverse-logarithmic corrections, to the mean-field scaling.

To account for these corrections, we first integrate the flow equation for \(u(\ell)\).
This gives the asymptotic running coupling
\begin{align*}
u(\ell)
=
\frac{u(0)}
{1+\frac{3}{2}u(0)\ell
-\frac{17}{9}u(0)\ln\!\left(
1+\frac{\frac{3}{2}u(0)}{1-\frac{17}{9}u(0)}\,\ell
\right)},
\end{align*}
Inserting \(u(\ell)\) into the flow equation for \(r(\ell)\) and integrating the RG flow up to the stopping scale \(\ell_0\), defined by \(r(\ell_0)\sim \mathcal{O}(1)\), identifies the point at which the system exits the critical regime.
This stopping scale determines the correlation length via \(\xi_{\mathrm{GS}}\sim e^{\ell_0}\).
This procedure yields a refined scaling form for the correlation length as a function of $r$,
\begin{align*}
    \xi_{\textrm{GS}} \propto r^{-1/2} &\left[\ln\left(\frac{r_0}{r}\right)\right]^{1/6}\\
    &\ \ \ \ \ \ \times\left[1 + \frac{\mathcal{G}_\mu(\ln\left(\frac{r_0}{r}\right))}{\ln\left(\frac{r_0}{r}\right)} 
    + \mathcal{O}\left(\frac{\ln\ln\left(\frac{r_0}{r}\right)}{\left[\ln\left(\frac{r_0}{r}\right)\right]^2}\right)\right],
\end{align*}
where \(r_0\) is a nonuniversal scale and the function
\(\mathcal{G}_\mu(x) = \frac{5}{27} + \frac{1}{3\mu} - \frac{34}{81}\ln\left(1 + \frac{\mu}{3}x\right)
\)
captures the sub-leading corrections.

To determine the freeze-out scaling, we evaluate the equilibrium scaling along the ramp, \(r=\hat t/\tau_q\), and impose the sonic-horizon condition \(\xi_{\textrm{GS}}(\hat t)\sim c\,\hat t\). 
Rather than inverting the closed form of \(\xi_{\textrm{GS}}(r)\) directly, we derive the freeze-out scaling from the underlying implicit RG relation between \(r\) and the stopping scale \(\ell_0=\ln\xi_{\textrm{GS}}\), which is more convenient for recursively extracting the logarithmic corrections. This yields the refined scaling law
\begin{align*}
    \hat{t} \approx \tau_q^{1/3} [\ln (C\tau_q)]^{1/9}
    \left( 1 + \frac{\mathcal{G}_\mu(\ln C\tau_q)}{\ln (C\tau_q)} \right),
\end{align*}
where the same function \(\mathcal{G}_\mu(x)\) captures the sub-leading corrections.

\section{Numerical procedure for the fitting}
\label{app:collapse}
In this Appendix, we describe the numerical fitting procedure used to determine the nonuniversal constants \(C\) and \(\mu\) entering the freeze-out length in the scaling collapse of the correlation function shown in Fig.~\ref{fig:figure2}.
The freeze-out length is taken from Eq.~\eqref{eq:refined_scaling} and written as
\begin{align*}
\hat{\xi}
=
A\,\tau_q^{1/3} [\ln(C\tau_q)]^{1/9}
\left(
1+\frac{\mathcal{G}_\mu(\ln(C\tau_q))}{\ln(C\tau_q)}
\right),
\end{align*}
where \(A\) is a nonuniversal prefactor.

To determine the fitting parameters, we minimize the root mean squared deviation of the rescaled correlation data from the expected exponential scaling form. The corresponding cost function is
\begin{align*}
\chi(A,C,\mu,K) = \sqrt{ \frac{1}{N_{\text{data}}} \sum_i \left[ \hat{\xi}_i^2\, C(R_i) - K \exp(-R_i/\hat{\xi}_i) \right]^2 }
\end{align*}
where the sum runs over all $N_{\text{data}}$ data points included in the collapse analysis, and \(C(R_i)\) denotes the measured correlation function.

Here, \(A\) and \(K\) account for overall nonuniversal factors in the freeze-out length and the scaling function, whereas \(C\) and \(\mu\) control the logarithmic structure of the refined scaling form.
The quality of the collapse is therefore governed by \(C\) and \(\mu\).
Even without introducing \(A\) and \(K\), the collapsed correlation function already exhibits an exponential form.
The role of \(A\) and \(K\) is only to absorb nonuniversal prefactors, so that a form \(K e^{-A R/\hat{\xi}}\) can be rewritten in the form \(K e^{-R/\hat{\xi}}\).

We perform the minimization using the \texttt{Optim.jl} package~\cite{mogensen2018optim}, employing Newton's method with a trust-region algorithm.
This yields the optimal parameter values \(A=0.34673, \mu=0.19376, C=81.18926, K=0.21475,\) with a minimum cost \(\chi_{\min}=8.84\times10^{-5}.\)

\bibliography{References}

\end{document}